\documentclass[preprint]{aastex}

\newcommand{\Msun}{M$_{\odot}$}

\newcommand{\teff}{T$_{\rm eff}$}

\received{15 February 2000}
\revised{12 July 2000}
\accepted{25 July 2000}

\shortauthors{Reed et al.}
\shorttitle{A Test of Asteroseismology}

\begin{document}

\title{PG2131+066: A Test of Pre-White Dwarf Asteroseismology}

\author{M. D. Reed \altaffilmark{1,3}}
\author{Steven D. Kawaler \altaffilmark{1,4}}
\author{M. Sean O'Brien \altaffilmark{1,2,5}}
\altaffiltext{1}{Department of Physics and Astronomy, Iowa State University, 
Ames, IA  50011  USA}
\altaffiltext{2}{Space Telescope Science Institute, Baltimore, MD 27701 USA}
\altaffiltext{3} {mreed@iastate.edu}
\altaffiltext{4} {sdk@iastate.edu}
\altaffiltext{5} {obrien@stsci.edu}

\begin{abstract}

PG~2131+066 is a composite--spectrum binary with a hot pulsating PG~1159-type
pre--white dwarf and an early M-type main sequence star.  Analysis of Whole
Earth Telescope observations of the pulsating pre-white dwarf component
provided an asteroseismological determination of its mass, luminosity, and
effective temperature.  These determinations allowed Kawaler et al.  (1995)
to determine the distance to this star.  In this paper, we refine the
asteroseismological distance determination, and confirm the distance by an
independent measurement to the system via the spectroscopic parallax of the M
star.  PG~2131+066 was observed by the HST using the original PC in September
1993.  Exposures with filters F785LP and F555W both showed the companion at a
distance of 0.3 arc seconds.  Photometry of the images provides an apparent
magnitude for the main sequence companion of $m_v=18.97\,\pm \, 0.15$, from
which we find a distance of $ 560\,^{+200}_{-134}$ pc.  We also recalculated
the asteroseismological distance to the pre--white dwarf using updated models
and new spectroscopic constraints from UV spectra.  The new seismological
distance is $668 ^{+78}_{-83}$ parsecs, in satisfactory agreement with the distance
of the secondary star.  These results suggest that this is indeed a 
physical binary,  and that the seismological distance determination 
technique may be the best way to determine the distance to the pulsating 
hot pre-white dwarf stars. 

\end{abstract}

\section{Introduction}

Asteroseismology of pre-white dwarfs (PWDs) provides fresh insights 
into their interior structure and bulk properties.  However, because 
of their extreme temperatures and surface gravities, spectroscopic 
verification of seismic results has been limited.  An
independent distance determination would provide a comparison between 
model and observed luminosities, providing a more confident 
interpretation of seismic and spectroscopic results.  Unfortunately, 
none of the GW~Vir type stars are close enough to obtain a distance 
via parallax.  However, PG~2131+066 (hereafter referred to as 
PG~2131) has a red main sequence companion, which provides an independent
test of its distance.

PG~2131 is a member of the PG~1159 spectral class of stars (the pulsating
PG~1159 stars are referred to as GW~Vir stars).  These hot, ($T_{\rm
eff}\approx 100,000K$) luminous ($L\approx 10-1000L_{\odot}$) stars represent
a link in evolution between planetary nebula nuclei and white dwarfs.  As
such, by understanding their interior conditions, we provide constraints on
both AGB and white dwarf stars.

Spectroscopy by Wesemael et al. (1985) revealed a red component which they
proposed as originating from a companion star (hereafter referred to as the
secondary star).  Wesemael et al.  (1985) deconvolved their spectrum into its
constituent components by using a Rayleigh-Jeans approximation to describe the
red side of the primary star spectrum.  The remaining flux was attributed to
the secondary, to which they assigned a spectral type of K5-M0 V, and
synthesized a $V$ magnitude.  From this, they determined a distance to the
secondary of $d=1047^{+1000}_{-500}$ pc.

PG~2131 was discovered to be a variable star by Bond et al. (1984).
Pulsations of PG~2131 were investigated in 1992 by the Whole Earth Telescope
(WET) collaboration.  Seismological analysis by Kawaler et al. (1995) of the
WET data, combined with the best available spectroscopic temperature of \teff
$= 80,000\,\pm\, 10,000$ K (Werner \emph{et al.} 1991) implied a luminosity
of log($L/L_{\odot}$)$=\,1.0\,\pm\, 0.2$.  Adopting a crude bolometric
correction, Kawaler et al.  (1995) derived a $V$ magnitude of 16.6, which
provided a ``seismic'' distance to the primary of $470\, ^{+180}_{-130}$ pc.

PG~2131 provides an important way to ``calibrate'' seismic results and
confirm (or challenge) the reliability of the procedures by which information
is extracted from the stellar pulsations.  The asteroseismological distance
is the end inference of seismic analysis (depending on the other parameters
such as stellar mass, luminosity, temperature etc. determined through earlier
steps). Independent testing of this procedure is required to establish the
reliability of this distance determination technique.  Such a confirmation has
been made for a cooler pulsating white dwarf, GD~358 (Bradley \& Winget 1994)
via trigonometric parallax, but an independent test of the distance
determination of a pulsating GW~Vir star is vital.

Though the two distances quoted above are within their rather large error
estimates we sought to improve both estimates using the best available
observations, atmospheric models, and seismic models.  Bond and colleagues
imaged PG~2131 in 1993 with the first-generation Planetary Camera on HST.
Even with the point-spread function problems, they resolved the binary into
its two components, and reported concordance between the distance determined
using the red companion and the asteroseismic distance from Kawaler et al.
(1995) (Bond, private communication).  In this paper, we report on our
photometric analysis of two HST frames that resolve the binary, compare the
photometric colors with models of both stars, and reevaluate the
seismological analysis.  Section 2 describes the HST photometry and our
analysis.  Section 3 uses the photometric determinations to compare with
stellar atmosphere models and obtain spectroscopic parallax distances.  In
Section 4, we review the seismological analysis using updated parameters for
the pre--white dwarf component.  We compare the results and conclude in
Section 5.

\section{Photometry}

In 1993, HST obtained images of PG~2131 through the F555W and F785LP filters
with WFPC1.  These images were originally obtained by Howard Bond as part of
a snapshot project to search for binary PN central stars.  These images
confirm the binary nature of PG~2131.  As expected, the images show a visual
companion that is as extremely red as the primary star is blue.  The
separation of the pair is 0.3", at a position angle of 21 degrees.

We used the IRAF package PHOT to obtain aperture photometry on the stellar
cores of both images.  To get relative photometry, we chose an aperture that
included the stellar core but excluded the first diffraction ring.  The small
angular separation of 0.3 arc seconds results in the core counts being
polluted by their neighbor.  This is particularly severe in the F555W image,
where nearly half of the secondary core counts can be attributed to the
primary.  To remove this contamination, we took advantage of the symmetry of
the PSF (see Figure 1).  We measured the counts on the opposite side of each
star from its companion, and subtracted this from the core counts of the
other star; the apertures are illustrated in Figure 1c.

\begin{figure}
\epsscale{0.7}
\plotone{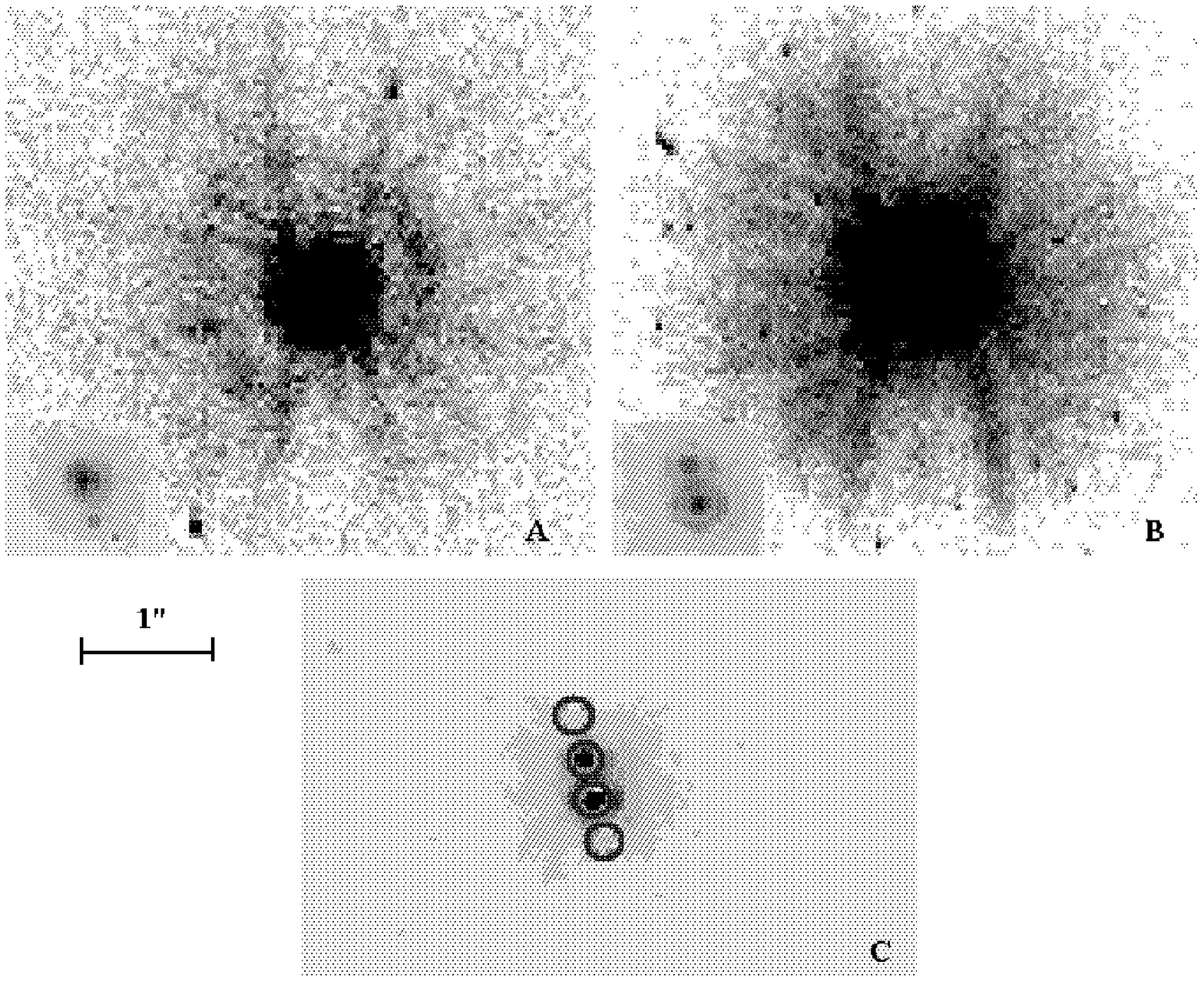}
\caption
{HST PC images of PG~2131.  Panel A) is the F555W image and panel B) is the 
F785LP image.  The inset to the lower left of these two panels shows, with 
a different stretch, the core of the image, revealing the two components.  
Panel c) is the F785LP Image of the PG~2131 system, with overlays showing
the apertures used. \label{fig01}}
\end{figure}

\placefigure{fig01}

With no other stars in the field, we needed to approximate absolute
photometry from this pair alone.  To do this, we accumulated all the
combined flux by using a large aperture ($\approx 2.4"$) reaching out
to points where the signal was indistinguishable from the sky
background.  Then we used the relative photometry to distribute the
total flux to each star.  This provided the ``absolute'' photometry given
in Table 1.

In the analysis that follows, the values we obtained were subject to
extensive Monte Carlo simulations.  These provided most--likely values
for all parameters from instrumental magnitudes through the final
distance determinations.  We simulated all photometric measurements
(i.e. counts and count rates) using Poisson statistics.  Uncertainties
in color corrections and other expressions were modeled as Gaussian
distributions with the standard deviations given in the sources of
these relations, as cited.  Errors were not propogated analytically,
but were determined throughout the sequence by the distribution of
values returned by the Monte Carlo trials.  The errors quoted are the
1 $\sigma$ errors derived from the Monte Carlo probability
distributions.

Using filter conversions from Holtzman et al. (1993), we iteratively 
determined the $V$ and $I$ magnitudes listed in Table 1 from the HST 
F555W and F785LP filter bandpasses.  It should be noted however that 
the Holtzman et al. (1993) conversions do not include stars as blue 
or red as the components of PG~2131, but it does contain the widest 
range available in the literature - from $V-I$ of 0 to approximately 1.7. 
We estimate conversion errors of 0.1 magnitudes for each filter.

\placetable{tab01}

If the secondary is a main sequence star, the $V-I$ color of the secondary
may be used to determine the absolute magnitude using an empirical
main-sequence $(M_{V},V-I)$ relation.  Several are available from the recent
literature.  Clemens et al.  (1998) used nearby stars, combined with
Hipparcos data, to determine such a relationship for the lower main
sequence.  Using their composite fit, we find that $M_{V}=10.01 \pm 0.68$,
yielding a distance of $560 ^{+200}_{-134}$ pc.  Error estimates combine the
uncertainty for the relationship quoted by Clemens et al. (1998) as well as
the photometric errors.  It is important to note here that if the companion
is not a main sequence star, but is a luminosity class III star of the same
color, it would have an absolute magnitude of -0.6, placing it over 78,000
pc. away!

We also obtained an absolute magnitude for the secondary 
from Figure 6 of Baraffe et al. (1998).  Using this figure, we find 
$M_{V}=9.68 \pm 0.1$.  With this absolute magnitude we 
calculate a distance for the secondary of $688\,\pm 65$ pc.  Even 
though this error is smaller, it is based on a region where the 
Baraffe et al. (1998) sample contains few stars, and therefore is more 
prone to systematic effects that are not in the error estimate.  (We 
use the value of $M_{V}=10.01 \pm 0.68$ from the Clemens et al. [1998]
relation for the rest of the paper.)

The only published ground--based photometry of PG~2131 is a magnitude given
by Green et al.  (1986) of 16.63; this entry in their tabulation is denoted
as uncertain.  This is a multichannel $V$ magnitude, which approximates
Johnson $V$.  For comparison, the combined V magnitude of the two stars on
the HST frame is $16.79 \pm 0.10$.  Given the large uncertainty in the
Green et al.  (1986) value, this represents consistency between the
ground-based photometry and the PC photometry.  We note that Bond et
al. (2000) have obtained newer ground-based photometry for comparison with 
the HST images.

\section{Model Fitting}

Given the $V-I$ color and distance for the secondary, we were 
ready to determine if the pair was an optical binary or a physical 
binary.  To do this, we used the data for the secondary to calibrate
models to the HST WFPC1 system. Then we determined the distance for the
primary by convolving models with the now calibrated system where the only
free parameter is distance.

Starting from the $V-I$ vs $T_{\rm eff}$ relationship of Monet et al. (1992),
we found \teff$ =3600 \pm 250$K for the secondary.  Next we processed a grid
of Kurucz models\footnote{The models included in our grid ranged from models
corresponding to K5~V stars to M5~V stars.} from the tabulation by Lejeune et
al. (1997) for comparison with our derived $V-I$ color and chose a model with
$T_{\rm eff}=3500K$ and $\log g=4.5$ as the best match; in good agreement
with our Monet et al. (1992) derived temperature.  

To calibrate our synthetic HST system, we first adopted stellar radii
for the two stars.  For the secondary, this was a radius appropriate
for an M2~V star, while the radius used for the primary is
$0.0182R_{\odot}$.  We convolved our M2~V model with the WFPC1 system
throughputs for the two filters, assuming the distance determined in
Section 2.  Comparing this flux with the observed flux provided the
appropriate zero point so the model magnitudes match the observed
magnitudes for the secondary.  With these zero points, we "observed"
models for the primary, adjusting the distance until our synthetic
magnitudes matched the HST observations.  Using this method, we
derived a distance of $681\,^{+170}_{-137}$ pc for the primary.
	
\placetable{tab02}

Though Paunzen et al. (1998) claim evidence for a 
possible third component in this system, the HST photometry shows that 
such a close companion to the white dwarf would produce much redder 
colors than are observed for it.  The spectroscopic signature of 
H$\alpha$ most likely comes from the M2V star visible on the HST 
images - this is reasonable considering the relative youth of this 
system (the ``newly minted'' white dwarf started out as a star of 
about 2.3 $M_{\odot}$ about $7.5 \times 10^{8}$ years ago).

\section{Seismological Analysis}

The primary of PG~2131 is a member of the GW~Vir class of pulsating 
pre-white dwarfs.  These stars are multiperiodic nonradial $g$-mode 
pulsators.  The periods present allow asteroseismological 
determination of bulk stellar properties such as mass and rotation 
rate.  In some cases, the period spectrum is rich enough to allow 
determination of the internal structural parameters such as the 
thickness of various composition layers.  Perhaps the best example of this
procedure is that shown for PG~1159 by Kawaler \& Bradley (1994).

Knowledge of stellar properties from other sources (such as spectroscopy or
broadband photometry) frequently provides key input in the seismic analysis.
For example, a given set of pulsation periods can be matched by a model of a
given mass (within a limited mass range) at discrete values of \teff .
Figure 2 shows the least-squares fit between model periods and the observed
periods in PG~2131 as a function of $T_{\rm eff}$  for four different stellar
masses.  Clearly, the fit is temperature dependent within a given mass, but
each shows a pronounced minimum at a well-defined effective temperature.  To
see the quality of the fit in a different way, we present Figure 3.  This
figure shows the observed periods of PG~2131 as dark horizontal lines;
periods of successive overtones of three models (0.60 \Msun, 0.61 \Msun, and
0.62 \Msun) are shown as thinner lines.  The intersection of the model
periods with the observed are marked as points; a perfect fit would be a
precise vertical alignment of points at the temperature of the best model.
The departures from a perfect fit could be the result of mode trapping by a
subsurface composition transition, as discussed in Kawaler \& Bradley (1994).
We did not make an effort to match these small departures with the models,
but clearly these departures could be reduced with such efforts without
affecting the \teff $\,$ fit significantly.

\begin{figure}
\epsscale{0.7}
\plotone{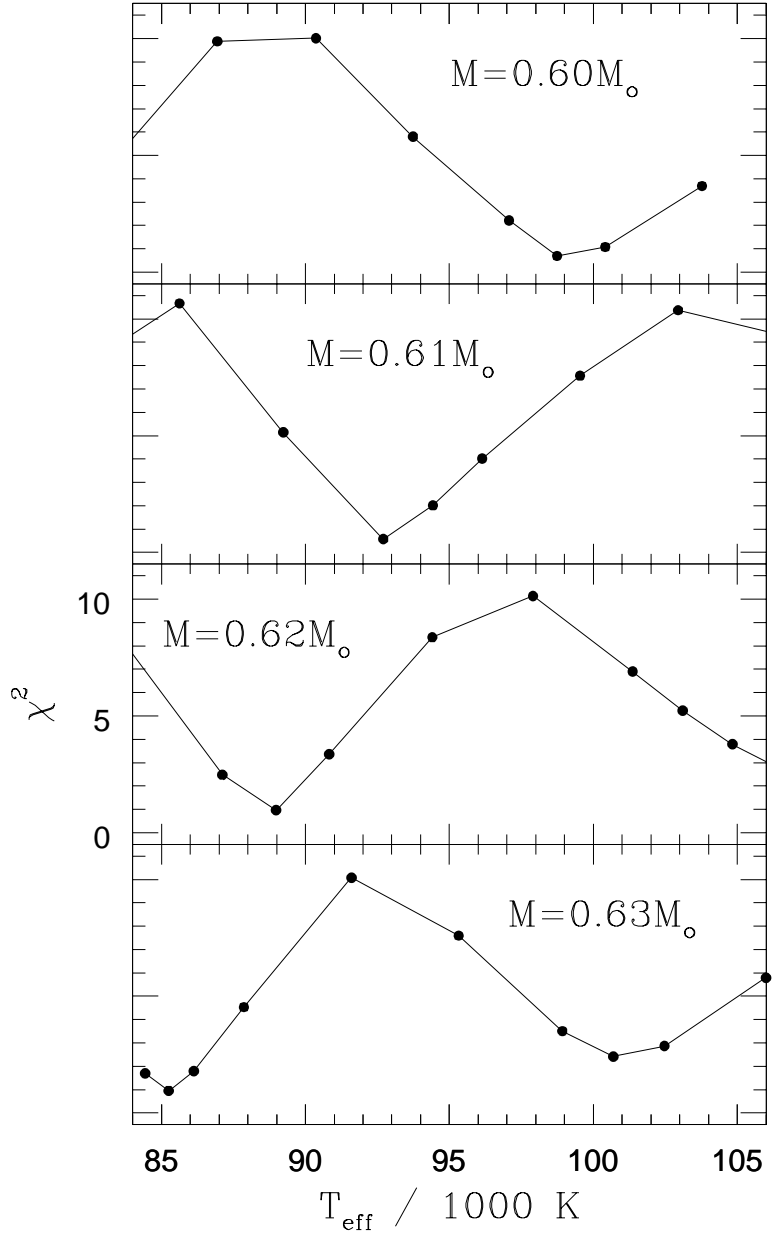}
\caption
{Comparison between the observed periods of PG~2131 and model 
periods.  The $\chi^{2}$ difference between the observed and model 
periods is plotted against the effective temperature of evolutionary 
models of the masses indicated.  Pronounced minima occur at 
well-defined effective temperatures. \label{fig02}}
\end{figure}

\placefigure{fig02}

\begin{figure}
\epsscale{0.7}
\plotone{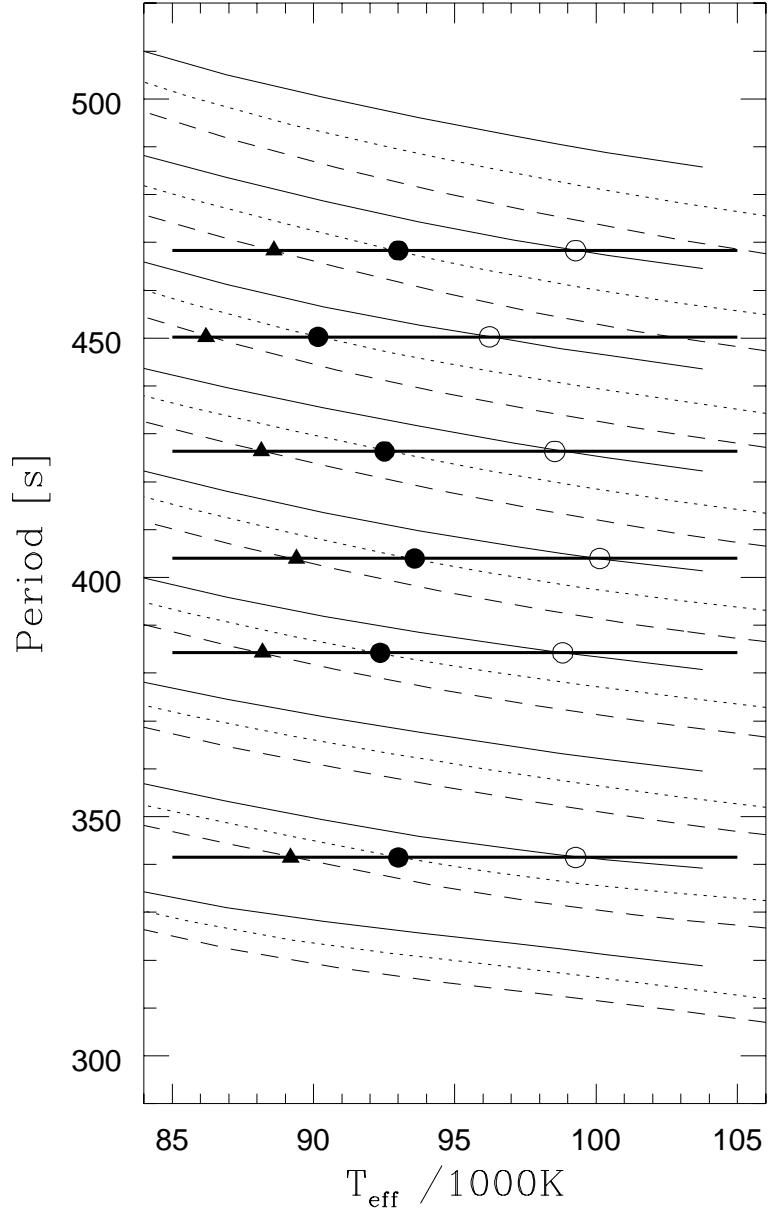}
\caption[pulsetab.eps]
{Periods of PG~2131 compared with model periods.  The thick horizontal 
lines represent the observed periods, while thinner lines are model 
periods for 0.60 \Msun (solid), 0.61 \Msun (dashed) and 0.62 \Msun 
(dotted).  Points of intersection are highlighted; a perfect-fit model 
would produce a vertically aligned set of highlighted points 
\label{fig03}}
\end{figure}

\placefigure{fig03}

Considering the range of possible masses, as limited by the 
spectroscopic temperature determination and its uncertainty, the 
period--matching constraint results in a relation between bolometric 
luminosity and effective temperature (as shown by the dashed line in 
Figure 4).  In the case of PG~2131, a specific value of $T_{\rm eff}$, 
combined with the requirement of matching the observed pulsation 
period, breaks the degeneracy of the models and isolates a model at a 
given mass and luminosity.

\begin{figure}
\epsscale{0.8}
\plotone{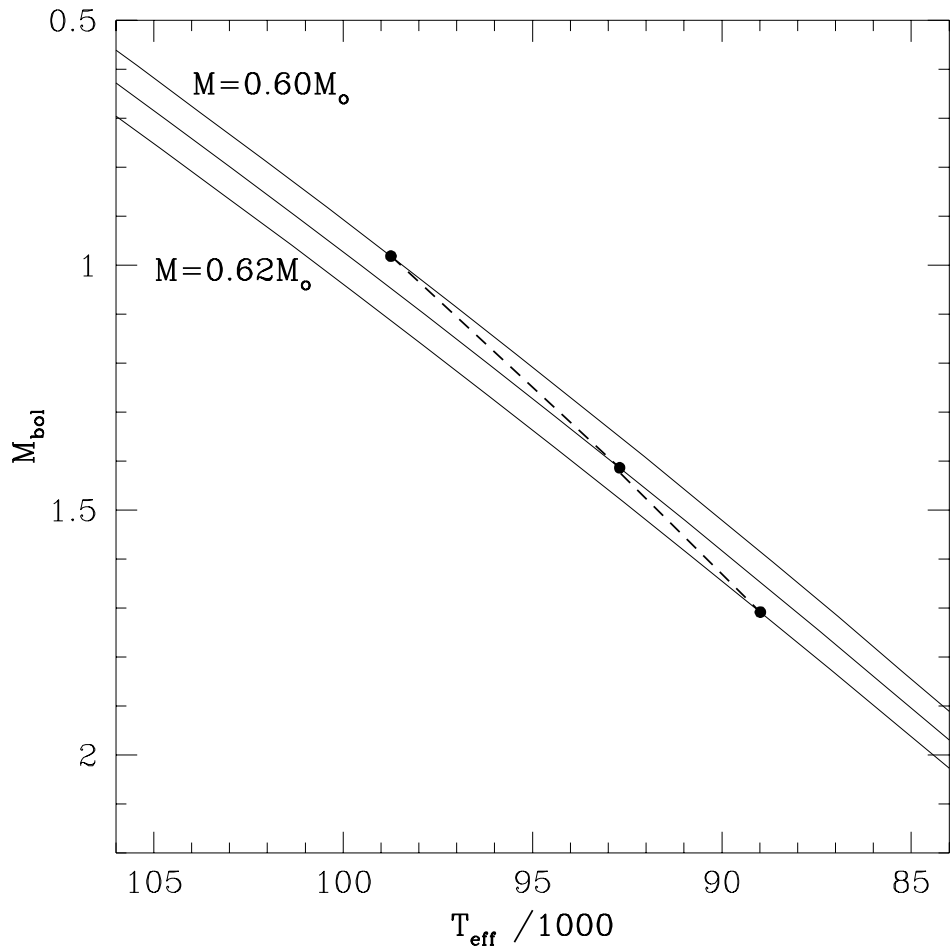}
\caption[hrd2131.eps]
{A theoretical H--R diagram for models of PG~2131.  The dashed line 
connects models that provide a good fit to the observed periods of 
PG~2131 as a function of temperature and which lie on the evolutionary 
tracks.  Thus the dashed line represents a period-luminosity relation 
for PG~2131. \label{fig04}}
\end{figure}

\placefigure{fig04}

Kawaler et al. (1995) used this procedure, along with the 
then--current spectroscopic temperature of 80,000K, to determine the 
mass and luminosity of PG~2131, and therefore its seismic distance.  
Later, Dreizler \& Heber (1998) revised the temperature upward to 
95,000K.  We used the revised temperature, and the models shown in Figures 
2-4, to recalculate the parameters of PG~2131.  Using Dreizler's 
atmospheric models, we calculated bolometric corrections to determine 
the ``seismic'' distance to PG~2131 of $668\,^{+79}_{-83}$ pc.  Most of 
the uncertainty in the seismic distance lies in the bolometric 
correction.  Table 2 shows the properties of the best-fit seismic 
model of PG~2131.

\section{Summary}

Though the HST images show the 0.3" separation between the two stars, 
obtaining accurate photometry with these images requires separating 
the flux into its constituent components.  We have used aperture 
photometry on the stellar cores and the entire field to obtain 
apparent magnitudes.  From the photometry, and a transformation to 
$V-I$ colors, we deduced that the secondary is an $M2\pm 1$ V star and 
determined it's distance to be $ 560\,^{+200}_{-134}$ pc.  We used 
this data to calibrate models of the WFPC1 system for use with stellar 
models.

By convolving atmosphere models for the primary with a synthetic WFPC1
system, we were able to obtain a reasonable match to the observed magnitudes
and colors of the primary star.  The best model fit is a GW~Vir type star at
an effective temperature of 95,000K.  For the radius of a pre--white dwarf
star, the observed magnitude yields a distance of $681 ^{+170}_{-137}$
parsecs, matching the distance determined using the secondary to within the
uncertainties.

A re-analysis of seismic data also indicate that the best fit for the 
primary is a PG~1159 type star at an effective temperature of 95,000K. 
However, seismic analysis also provides constraints for mass, radius 
and luminosity.  Applying a theoretical bolometric correction to the 
primary, the luminosity can be used to determine the distance.  We 
find that the seismic distance is $668\,^{+79}_{-83}$ pc, entirely 
consistent with the photometry.

Table 3 summarizes the distance estimates for PG~2131.
We conclude that the system is a physical binary.  Taking a weighted mean 
of the seismic distance to the primary and the photometric distance to the 
secondary provides a distance to the system of $632\,^{+150}_{-111}$ pc.

\placetable{tab03}

Though the photometric and seismic results are comforting, the large
errors are still problematic.  The error budget for the spectroscopic
parallax is dominated by the $M_{V}$--$(V-I)$ relation.  It is not the
photometry that is at fault.  Our knowledge of the late main sequence
is lacking.  Even though Clemens et al.  (1998) used Hipparcos data to
derive the relation between $M_{V}$ and $V-I$, the absolute magnitude
has an uncertainty of 0.35 magnitudes.  Add to that an error of 0.1
magnitudes in the filter conversions, and the slim possibility remains
that this is not a physical pair.  A new image taken with WFPC2 would
be useful in two ways.  First, if any differential proper motion is
present, then that would show that the pair is not a physical pair. 
The new image would also greatly reduce the photometric errors, but we
would still depend on tighter constraints on the late main sequence to
refine the distance through spectroscopic parallax. 

Still, the determination of asteroseismological distances to white 
dwarfs represents the ``end product'' of the analysis pipeline --- 
thus any systematic or internal errors in the analysis will result in 
an erroneous distance.  The remarkable agreement between the 
asteroseismological distance to PG~2131+066 and the more traditional 
determinations, coupled with the smaller uncertainties, suggests that 
the asteroseismological techniques may be the most precise and 
reliable distances yet obtained for these stars.

\acknowledgments

The authors wish to thank Howard Bond for numerous discussions during the
course of this work.  Stefan Dreizler kindly provided details of his models
for PG~1159 stars.  We gratefully acknowledge the National Science
Foundation for support under the NSF Young Investigator Program (Grant
AST-9257049) to Iowa State University, and the NASA Astrophysics Theory
Program through award NAG5-4060 and NAG5-8352.   The HST Archive provided a
valuable resource for obtaining and reducing the images.

\newpage


\clearpage

\begin{deluxetable}{lcc}
\tablecaption{Observed magnitudes \label{tab01}}
\tablehead{
\colhead{Magnitude/Color} & \colhead{Primary} & \colhead{Secondary}}
\startdata
$F555W$           & $16.84\,\pm\, 0.10$ & $19.07\,\pm\, 0.12$ \\
$F785LP$          & $17.38\,\pm\, 0.11$ & $16.54\,\pm\, 0.10$ \\
$F555W - F785LP$  & $-0.54\,\pm\, 0.15$ & $ 2.53\,\pm\, 0.16$ \\
$V$               & $16.86\,\pm\, 0.14$ & $18.97\,\pm\, 0.15$ \\
$I$               & $17.30\,\pm\, 0.15$ & $16.85\,\pm\, 0.13$ \\
$V-I$             & $-0.46\,\pm\, 0.20$ & $ 2.10\,\pm\, 0.19$ \\ 
\enddata
\end{deluxetable}


\begin{deluxetable}{cccc}
\tablecaption{Seismic results \label{tab02}}
\tablehead{
\colhead{Mass [M$_{\odot}$} &
\colhead{Luminosity [L$_{\odot}$]} &
\colhead{Radius [R$_{\odot}$]} &
\colhead{M$_{\rm v}$}}
\startdata
$0.608 \pm 0.011$ &
$26^{+10}_{-8}$  &
$0.0186 \pm 0.0012$ &
$7.69 ^{+0.25}_{-0.18}$\\
\enddata
\end{deluxetable}


\begin{deluxetable}{cccc}
\tablecaption{Distance determinations for PG~2131\label{tab03}}

\tablehead{\colhead{Method} & \colhead{Star} & \colhead {Reference} 
  & \colhead{Distance [pc]}}
\startdata
     spectrum   & secondary & Wesemael et al. (1985) & 1047 $^{+1000}_{-500}$\\
     seismology &   primary & Kawaler et al. (1995)  &  470 $^{+180}_{-130}$\\
 & & & \\
spect. parallax & secondary & this work              & 560  $^{+200}_{-134}$ \\
spectrum fitting&   primary & this work              & 681 $ ^{+170}_{-137}$ \\
 & & & \\
seismology      &   primary & this work              & 668 $ ^{+78}_{-83}$ \\
\enddata
\end{deluxetable}

\end{document}